\documentclass[twocolumn,superscriptaddress,showpacs,prl,aps,amsmath,amssymb]{revtex4}

\usepackage{graphicx}
\usepackage{bm}
\usepackage{tabularx}

\begin{document}

\title{
Spin nematic order
in multiple-spin exchange models on the triangular lattice
}
\author{Tsutomu Momoi}
\affiliation{Condensed Matter Theory Laboratory, RIKEN,
Wako, Saitama 351-0198, Japan}
\author{Philippe Sindzingre}
\affiliation{Laboratoire de Physique Th\'eorique de la Mati\`ere Condens\'ee, UMR
7600 of CNRS,  \protect\mbox{Universit\'e P. et M. Curie, case 121,
4 Place Jussieu, 75252 Paris Cedex, France}}
\author{Kenn Kubo}
\affiliation{Department of Physics and Mathematics, Aoyama Gakuin University,
5-10-1 Fuchinobe, Sagamihara, Kanagawa 229-8558, Japan}

\date{\today}

\begin{abstract}
We figure out that the ground state of a multiple-spin exchange
model applicable
to thin films of solid $^3$He
possesses an octahedral spin nematic order.
In the presence of magnetic field, it is deformed into
an antiferro-quadrupolar order in the perpendicular spin plane,
in which lattice $Z_3$ rotational symmetry is also broken.
Furthermore, this system shows a narrow magnetization plateau
at half, $m/m_{\rm sat}=1/2$,
which resembles recent magnetization
measurement [H.\ Nema {\it et al.}, Phys.\ Rev.\ Lett.\
\textbf{102}, 075301 (2009)].
\end{abstract}

\pacs{
75.10.Jm,
75.40.Cx
}
\maketitle

Frustrated quantum antiferromagnets have
long been a subject of active research~\cite{MisguichL},
since Anderson\cite{Anderson} suggested that a spin-1/2 Heisenberg antiferromagnet
on the triangular lattice would have a gapless spin liquid ground state named as
resonating-valence-bond (RVB) state.
Recent experimental studies of quasi-two-dimensional compounds,
such as solid $^3$He films absorbed on graphite~\cite{greywall89,IshidaMYF,MasutomiKI},
the organic Mott insulator~\cite{Shimizu}
$\kappa$-(BEDT-TTF)$_2$Cu$_2$(CN)$_3$ and
the transition metal chloride Cs$_2$CuCl$_4$~\cite{Coldea},
have further prompted theoretical research of
quantum spin liquid and competing exotic orders in
triangular lattice
antiferromagnets~\cite{Balents,KuboM,MomoiKN,MisguichBLW,MisguichLBW,MomoiSK,MomoiSS}.

Among these,
solid $^3$He films offer a perfect realization of a spin-1/2
triangular lattice. They contain a unique character in spin exchange interactions.
The nearest-neighbor interaction is {\it ferromagnetic} (FM) and
competes with antiferromagnetic (AF) multiple-spin cyclic exchange~\cite{roger90,CollinTHRBBG}.
As such,
$^3$He on graphite belongs to a new class of quasi-two dimensional
systems to exhibit ``frustrated
ferromagnetism''~\cite{dresden,kageyama}.
It is also unique in the possibility of tuning the ratio of the
competing interactions continuously by varying the density of
$^3$He atoms.
An additional strong motivation for understanding this
system comes from the fact that, for a range of densities bordering
on ferromagnetism, spins in solid $^3$He films
exhibit anomalous double-peak structure in
specific heat~\cite{IshidaMYF} with gapless
excitations~\cite{MasutomiKI}.

The effective Hamiltonian for nuclear magnetism of $^3$He
thin films
is given by the $S=1/2$ multiple-spin exchange (MSE) model
on the triangular lattice~\cite{roger90}.
The Hamiltonian containing up to six-spin exchanges and
applied field is written as
\begin{align}
& H_{eff} = J \sum_{\langle i,j\rangle} P_2
+ J_4 \sum_{\begin{picture}(26,12)(-2,-1)
        \put (0,0) {\line (1,0) {12}}
        \put (6,10) {\line (1,0) {12}}
        \put (0,0) {\line (3,5) {6}}
        \put (12,0) {\line (3,5) {6}}
        \put (6,10) {\circle*{3}}
        \put (18,10) {\circle*{3}}
        \put (0,0) {\circle*{3}}
        \put (12,0) {\circle*{3}}
        \end{picture}}
(P_4+P_4^{-1}) \nonumber\\
&- J_5\sum_{\begin{picture}(26,12)(-2,-1)
        \put (0,0) {\line (1,0) {24}}
        \put (6,10) {\line (1,0) {12}}
        \put (0,0) {\line (3,5) {6}}
        \put (18,10) {\line (3,-5) {6}}
        \put (6,10) {\circle*{3}}
        \put (18,10) {\circle*{3}}
        \put (0,0) {\circle*{3}}
        \put (12,0) {\circle*{3}}
        \put (24,0) {\circle*{3}}
\end{picture}} (P_5+P_5^{-1})
+ J_6\sum_{
\begin{picture}(26,22)(-2,-11)
        \put (6,10) {\line (1,0) {12}}
        \put (6,-10) {\line (1,0) {12}}
        \put (0,0) {\line (3,5) {6}}
        \put (0,0) {\line (3,-5) {6}}
        \put (18,10) {\line (3,-5) {6}}
        \put (18,-10) {\line (3,5) {6}}
        \put (6,10) {\circle*{3}}
        \put (6,-10) {\circle*{3}}
        \put (18,10) {\circle*{3}}
        \put (18,-10) {\circle*{3}}
        \put (0,0) {\circle*{3}}
        \put (12,0) {\circle*{2}}
        \put (24,0) {\circle*{3}}
\end{picture}}
(P_6 + P_6^{-1}) 
-h \sum_i S_i^z, \label{eq:H}
\end{align}
where $P_n$ denotes the cyclic permutation operator
of $n$ spins.
The summations in front of the permutation operators
run over all minimal $n$ spin clusters. The
first and the second summations, for example, run over all pairs of nearest
neighbors and all four-spin diamond clusters, respectively. In solid
${}^3$He on graphite, the effective two-spin coupling $J$ is
negative (FM) and the other couplings $J_n$ ($n=4,5,6$) are positive.
Experimental estimates suggest that two-spin and four-spin exchange interactions
are dominant, but five-spin and six-spin interactions are also
not small~\cite{CollinTHRBBG}.
We set $J=-2$ and the ratio $J_6/J_5=2$ throughout this letter.

Magnetism induced by multiple-spin exchange on the triangular lattice
has been studied mostly in the
$J$--$J_4$ model~\cite{KuboM,MomoiKN,MisguichBLW,MisguichLBW,MomoiSK,MomoiSS},
which contains two-spin
and four-spin exchanges.
When FM $J$ strongly competes with AF $J_4$, it was found that,
in a finite range bordering on ferromagnetism,
three magnon bound states become the most stable magnetic particles,
giving rise to an octupolar ordered phase
in applied magnetic field~\cite{MomoiSS}.
However the nature of the ground state
in the absence of magnetic field was not identified in this regime.
When four-spin exchange $J_4$ is dominant, exact diagonalization analysis
concluded a quantum disordered state with a large spin gap~\cite{MisguichBLW,MisguichLBW}.
The strong $J_4$ also supports a wide magnetization plateau
at half of saturation ($m/m_{sat}=1/2$), which originates from the appearance
of four-sublattice spin density wave with \emph{uuud}
structure~\cite{KuboM,MisguichBLW,MomoiSK}.

\begin{figure}[tb]
\begin{center}
    \includegraphics[width=42mm]{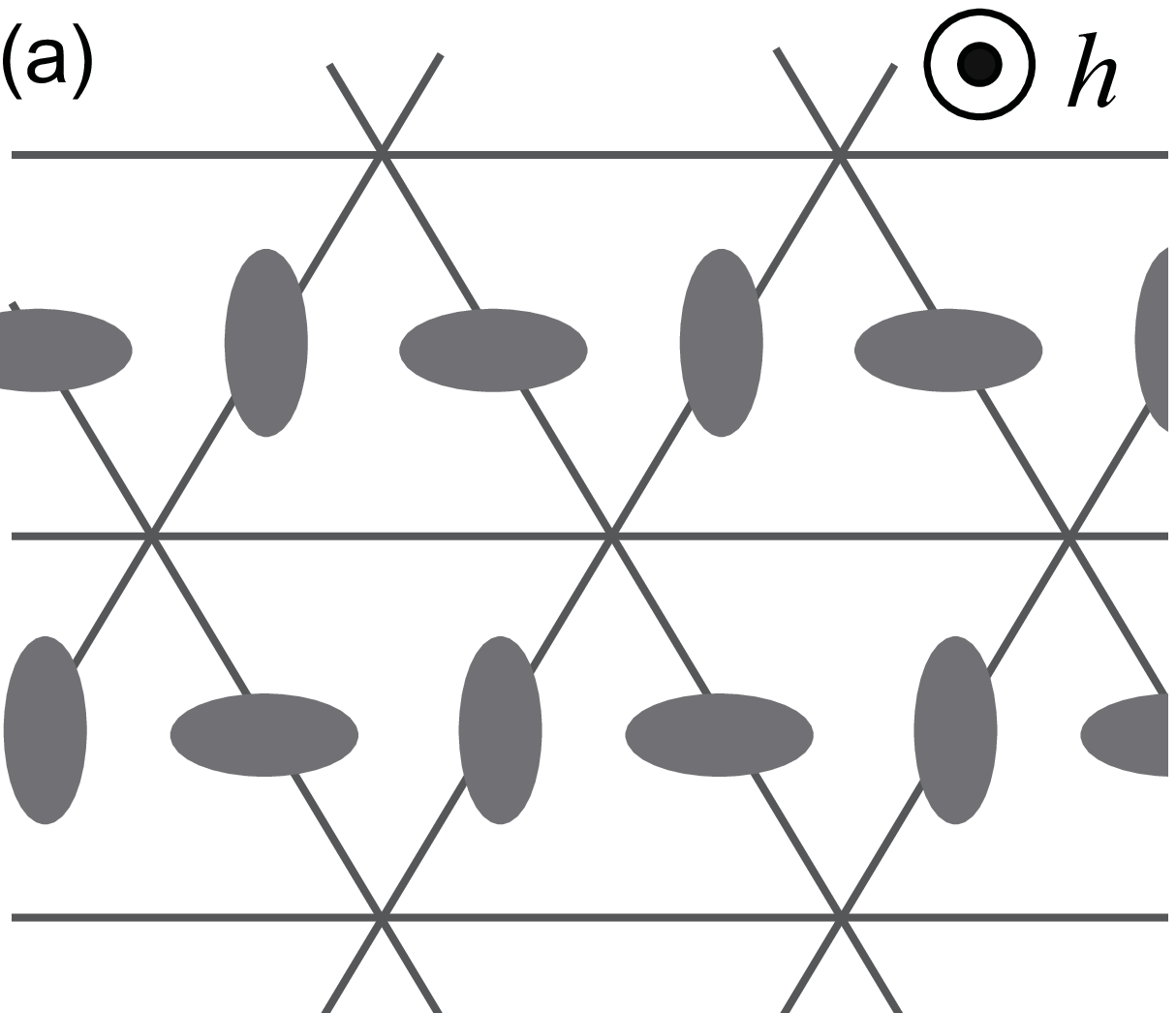}\hspace{1mm}
    \includegraphics[width=42mm]{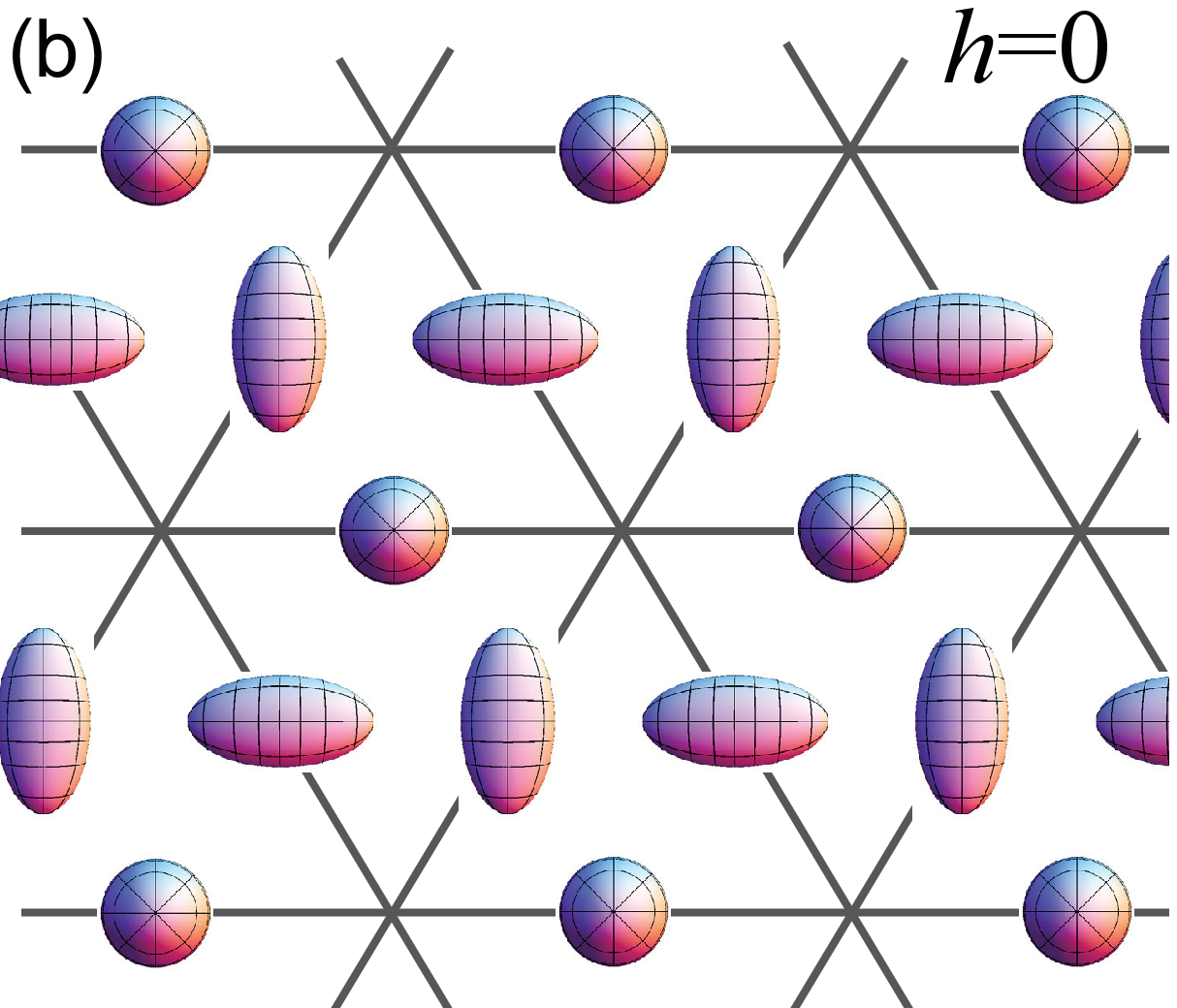}
\caption{Director configurations in spin
nematic order in $S=1/2$ MSE model on the triangular lattice.
Ellipses on bonds represent nematic-directors.
In the applied field (a), only spin transverse components are depicted and
there are also induced uniform dipole moments on sites.
At zero field (b), three types of director vectors are orthogonal to each other.
} \label{fig:d_config}
\end{center}
\end{figure}
In this letter, we study
$S=1/2$ MSE model
on the triangular lattice
containing up to six-spin exchange couplings [Eq.~(\ref{eq:H})],
which is directly applicable to solid $^3$He films.
Inclusion of these multiple spin exchanges helps us to identify
that the ground state is
an octahedral spin nematic state [Fig.~\ref{fig:d_config}(b)],
which has both liquid-like character and antiferro-quadrupolar orders on bonds.
In applied field, this spin state is continuously deformed into
a spin nematic state with antiferro-quadrupolar order
[Fig.~\ref{fig:d_config}(a)], where lattice $Z_3$ rotational symmetry is also broken.
This state is understood as condensation of $d+id$-wave two magnon bound states.

\begin{figure}[tb]
\begin{center}
    \includegraphics[width=82mm]{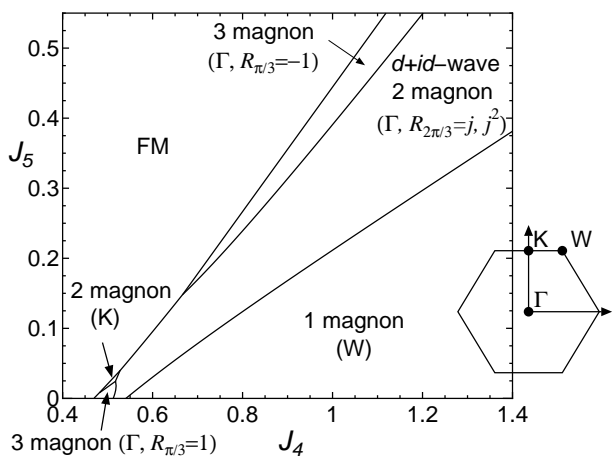}
\caption{
Magnon instabilities to the fully polarized state at the
saturation field. Parameters are set as $J=-2$ and $J_6=2 J_5$.
The wave vectors in the Brillouin zone ($\Gamma$, K, W, defined in the inset)
and some space symmetries are shown. FM denotes the stable ferromagnetic
phase.} \label{fig:Instability2J5}
\end{center}
\end{figure}
The occurrence of this spin state may indeed
be most easily understood from the channel  of magnon instability
at saturation field. As shown in Fig.\ \ref{fig:Instability2J5},
magnon pairs form stable bound states in a wide range
near the FM phase boundary.
As previously seen~\cite{Chubukov,MomoiS,ShannonMS},
condensation of bound magnon pairs leads to a spin nematic state~\cite{AndreevG},
which has
quadrupolar order
in the perpendicular component to the applied field and
breaks the U(1) symmetry of the Hamiltonian.
In the triangular lattice $J$--$J_4$ model (i.e. $J_5=J_6=0$),
instabilities given by two magnon bound states~\cite{MomoiS} and
three magnon bound states~\cite{MomoiSS} are competing.
Inclusion of five and six-spin exchange interaction
removes this competition, making the
$d_{x^2-y^2}+id_{xy}$-wave two magnon instability most dominant
in a wide parameter range near FM phase boundary,
as shown in Fig.~\ref{fig:Instability2J5}.
The $d+id$-wave magnon pairing operators are given by
$Q_{+}=\sum_i (S_i^-S_{i+e1}^- +jS_i^-S_{i+e2}^-
+j^2S_i^-S_{i+e1-e2}^-)$
and its complex conjugate $Q_{-}$,
where $j=\exp(i 2\pi/3)$ and ${\bm e}_1=(1,0)$, ${\bm e}_2=(1/2,\sqrt3/2)$.
The parameter set \cite{CollinTHRBBG} estimated in the 4/7 phase
of two-dimensional (2D) solid $^3$He
also belongs to this instability region.

The $d+id$-wave bound magnon pairs
$Q_{\pm}|{\rm FP}\rangle$,
where $|{\rm FP}\rangle$ denotes the fully polarized state,
are degenerate with chiral degrees of freedom as
the eigenvalues to the space rotation by $2\pi/3$ are $R_{2\pi/3}=j,j^2$.
In such a case, strong repulsion between different species of bosons
can induce density
imbalance between the condensates
of two species, breaking additional chiral $Z_2$ symmetry.

\begin{figure}[tb]
\begin{center}
    \includegraphics[width=85mm]{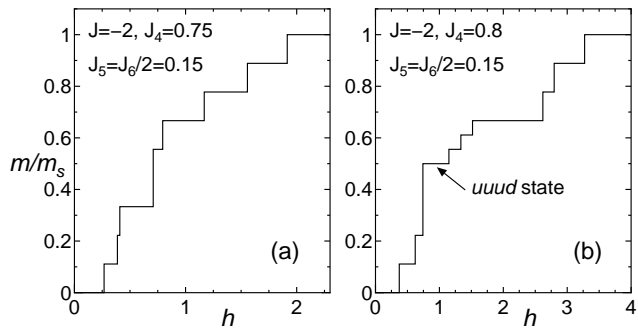}
\caption{
Magnetization process $m/m_{\rm s}$ of the MSE model for $N=36$ spin cluster.
The exchange parameters are
$J=-2$, $J_5=\frac12 J_6=0.15$ with (a) $J_4=0.75$ and
(b) $J_4=0.8$.
Jumps of $|\Delta m|=2$, corresponding to spin nematic
order, are clearly visible.
In (b), a narrow plateau structure appears at $m/m_{\rm s}= 1/2$.
} \label{fig:mag_pro}
\end{center}
\end{figure}

\begin{figure}[tb]
\begin{center}
    \includegraphics[width=80mm]{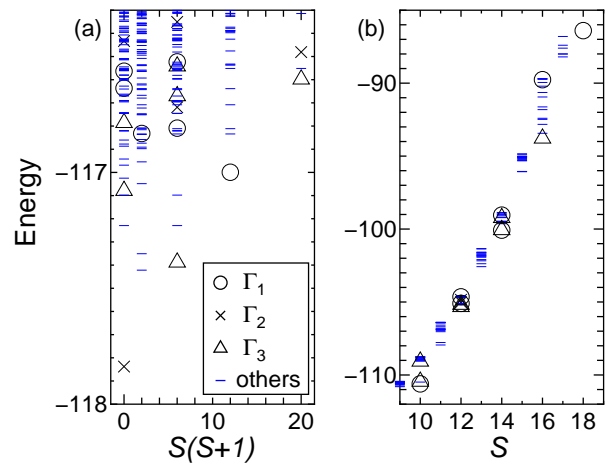}
\caption{(Color online) Energy spectrum of the MSE model for $N=36$ spin cluster
with (a) $J=-2$, $J_4=0.5$, $J_5=J_6=0$ near the singlet ($S=0$) and
(b) $J=-2$, $J_4=1$, $J_5=\frac12 J_6=0.3$
near the saturation ($S=18$).
The symbols $\Gamma_1$, $\Gamma_2$, and $\Gamma_3$ are defined in Table~\ref{table:IRP}.
} \label{fig:QDJS}
\end{center}
\end{figure}

To identify the nature of symmetry breaking,
we have performed numerical exact diagonalization
of Eq.~(\ref{eq:H}) for clusters
up to $N=36$ spins.
For parameter sets in the two-magnon instability regime,
the special stability
of lowest energy states in even $S$ sector
at large magnetization (high $S$ states), as seen in Fig.~\ref{fig:QDJS}(b),
signals the formation of bound magnon pairs with repulsive interactions,
which
leads to jumps by $|\Delta S|=2$ in the magnetization process
as shown Fig.~\ref{fig:mag_pro} and points to spin nematic ordering
in the perpendicular spins~\cite{ShannonMS}.

 From the analysis of irreducible representations (irreps) and
quasi-degeneracy of these low-lying states, we conclude to non-chiral
spin nematic ordering. 
A quasi three-fold degeneracy is observed
in low-lying states in the even spin $S$ sector
(for $S<m_{\rm s}-2$), instead
of the two-fold degeneracy expected for the chiral nematic.
It indicates that bound magnon pairs with different
chiralities condense with the same density and an additional
$Z_3$ symmetry is broken.

The order parameter of this non-chiral nematic state is identified as
\begin{equation}
{\cal O}_{U(1)}=Q_+ - Q_-= i\sqrt3 \sum_i ( S_i^-S_{i+{\bm e}_1}^- - S_i^-S_{i+{\bm e}_2}^-),
\end{equation}
by expansion of the coherent state $\exp[-\lambda{\cal O}_{U(1)}]|{\rm FP}\rangle$
into irreps, which gives the numerically observed irreps.
The required low-lying states for this order
are listed in Table~\ref{table:IRP}(a).
Note that space rotational $Z_3$ symmetry is broken in this order parameter, corresponding to
the choice of two bonds out of three.
A schematic figure of nematic-directors for this state is
depicted in Fig.~\ref{fig:d_config}(a).
The ground state manifold has spin $U(1)/Z_2$ symmetry and space $Z_3$ symmetry.

\begin{table}[t]
\newcolumntype{Y}{>{\centering\arraybackslash}X}%
\begin{tabularx}{85mm}{r|Y|Y|Y}
  \hline
\multicolumn{4}{l}{(a) $h>0$, $U(1)$ symmetric case} \\
  \hline
$m$ & $m_{\rm s}-2$ & $m_{\rm s}-4n$ & $m_{\rm s}-2(2n+1)$ \\
  \hline
Irreps & $\Gamma_3$ &  $\Gamma_1$, $\Gamma_3$
 & $\Gamma_2$, $\Gamma_3$ \\
 \hline
\end{tabularx}
\smallskip

\begin{tabularx}{85mm}{r|Y|Y|Y|Y|Y|c}
  \hline
\multicolumn{7}{l}{(b) $h=0$, $SU(2)$ symmetric case}\\
  \hline
$S$ & 0 & 2 & 3 & 4 & 5 & 6 \\
  \hline
Irreps & $\Gamma_2$ & $\Gamma_3$ & $\Gamma_1$ &  $\Gamma_2$, $\Gamma_3$ &
 $\Gamma_3$ & $\Gamma_1$, $\Gamma_2$, $\Gamma_3$ \\
 \hline
\end{tabularx}
\caption{\label{table:IRP} Irreps of low lying states in each total spin $S$
(magnetization $m$)
sector, needed for antiferro-quadrupolar ordering,
(a) in applied field  and (b) at zero field.
The symbols are defined as
$\Gamma_1\equiv(R_{2\pi/3}=1,\ R_\pi=1,\ \sigma=1)$,
$\Gamma_2\equiv(R_{2\pi/3}=1,\  R_\pi=1,\ \sigma=-1)$,
$\Gamma_3\equiv(R_{2\pi/3}=j,j^2,\ R_\pi=1)$, and all of them have
the wave vector ${\bm k}=(0,0)$.}
\end{table}

In the absence of magnetic field, $SU(2)$ symmetry is restored  and
the signature of $SU(2)$ symmetry breaking 
is the existence of an ``Anderson tower"
of quasidegenerate joint states (QDJS) belonging to different spin sectors
which form the $N=\infty$ ground state 
with energies scaled as $E_{QDJS}\sim \frac{S(S+1)}{N}$
and are well separated (at finite $N$) from the lowest magnon excitations
(scaled as $\sim 1/\sqrt{N}$).
The energy spectrum around $S=0$ is shown in Fig.~\ref{fig:QDJS}(a) for the
parameter set $J=-2$, $J_4=0.5$, $J_5=J_6=0$. A similar spectrum structure
is also found for finite $J_5$, for example in $J=-2$, $J_4=0.8$, $J_5=\frac12 J_6=0.15$.
The spin $S=1$ sector has high energy, which excludes conventional spin ordering.
Comparing irreps of the low energy states
with those expected for various possible spin structures, we
found that these states perfectly match with the Anderson tower
of a spin nematic state with three sets of orthogonal directors, 
whose order parameter is given by
\begin{equation}
{\cal O}_{SU(2)}=\sum_i \{ Q^{xx}_i({\bm e}_1)+Q^{yy}_i({\bm e}_2)
+Q^{zz}_i({\bm e}_1-{\bm e}_2) \},
\end{equation}
where
$Q^{\alpha\alpha}_i({\bm r})= S_i^\alpha S_{i+{\bm r}}^\alpha
-\frac13 \langle {\bm S}_i \cdot {\bm S}_{i+{\bm r}} \rangle$.

This order parameter has octahedral spin symmetry $O$,
if one combines spin transformations with the space group $C_{6v}$ of the triangular lattice.
To obtain irreps specific to this spin state, one may decompose the irreps $D^S$ of $SU(2)$
into irreps of $O$ in each spin $S$ sector~\cite{Hamermesh}
and remove irreps odd under $\pi$ rotation ($C_4^2$),
i.e.\ the three dimensional irreps $F_1$ and $F_2$, which should be absent
as the nematic order parameter is invariant under $C_4^2$ (since directors are headless).
This allows  projection only on $A_1$, $A_2$, and $E$.
For $N=36$ spins, $A_1$, $A_2$, and $E$ map onto
$\Gamma_1$, $\Gamma_2$, and $\Gamma_3$ of $C_{6v}$,
respectively ($\Gamma_\mu$ are defined in Table~\ref{table:IRP}).
The irreps for $N=36$ in each spin $S$ sector are shown in Table~\ref{table:IRP}(b),
which agrees well with symmetries of QDJSs found in Fig.\ \ref{fig:QDJS}(a).
The mapping of $A_1$ and $A_2$ to  $\Gamma_1$ and $\Gamma_2$ may interchange
depending on the number of spins $N$.
(Note that, only
the relative symmetries between QDJSs are important
for constructing symmetry broken states).
With this assignment, irreps around $S=0$ smoothly connect with those of low-lying
states determined
by the two-magnon instability at saturation field.
While the dynamics of the order parameter of a non-collinear N\'eel is that of a
symmetric top, which has $2S+1$ states in each total spin $S$ and 
magnetization $m$ sector,
here the ground state manifold has spin $SU(2)/D_2$ symmetry and so
the Anderson tower has  $S/2+1$ [$(S-1)/2$] states for  even (odd) $S$ sector, the same as
a symmetric top which are even under $\pi$ rotations
around the principal axes.  

In applied field, two orthogonal nematic directors become perpendicular to the field and
the other doesn't have any order in the perpendicular plane,
leading to the space $Z_3$ symmetry breaking.
The QDJSs around $S=0$ smoothly merge with the series of low-lying states at high magnetization.
Thus the octahedral spin nematic state at zero field is continuously deformed into
the spin nematic state under the field.

The order parameter ${\cal O}_{SU(2)}$ has the same spin symmetry as the
antiferro-quadrupolar (AFQ) state discussed
in the triangular lattice
$S=1$ bilinear-biquadratic model \cite{TsunetsuguA,LaeuchliMP}.
The irreps of QDJSs for the $S=1$ AFQ state, obtained from the decomposition
of the wave function~\cite{PencL}, are equivalent to the Anderson towers
of the $S=1/2$ octahedral spin nematic state at zero field,
through the mapping of the space groups.
The difference is that the quadrupolar moments exist on bonds in the $S=1/2$
spin nematic state, whereas on sites in the $S=1$ AFQ state.
In $S=1/2$ spin nematics, $S=1$ degrees of freedom are formed
on bonds
and moving around from bond to bond. The ground state is translationally
invariant at any field, i.e., it doesn't show any spin density wave,
which is also different from the $S=1$ system~\cite{LaeuchliMP}.
Thus, the $S=1/2$ spin nematic state  has a liquid-like
character~\cite{ShannonMS,ShindouM} and
cannot be decoupled into the site product state.
On the other hand, the $S=1$ AFQ state is formed by the product of the
quadrupolar moments localized on sites,
having three sublattice structure with the wave vector ${\bm k}=(4\pi/3,0)$.
%

The signature of spin nematic ordering in energy spectrum around $S=0$
was well detected
in a wide parameter range $J_5 \alt 0.2$ which
covers the $d+id$-wave two-magnon instability region at saturation field
slightly extending to the smaller $J_4$ regime. In most of the region,
the spin nematic state exists from zero magnetization to the saturation.
Near $J_5=J_6=0$ regime, three magnon instability induces an octupolar order
under magnetic field~\cite{MomoiSS},
but, at very low magnetization, the octahedral spin nematic order overcomes it,
as shown in Fig.\ \ref{fig:QDJS}(a).
For the $J_5>0.3$ range, there is no more clear separation of QDJSs and
also no signature of spin ordering.

\begin{figure}[tb]
  \centering
  \includegraphics[width=60mm]{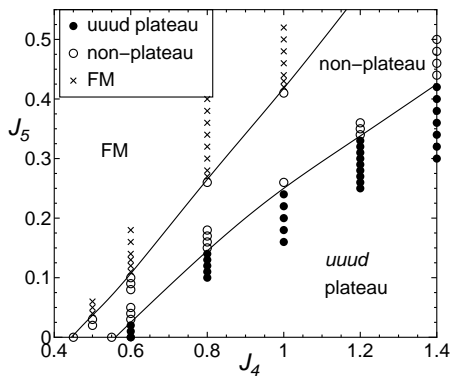}
\caption{
Phase diagram of the MSE model
in $m/m_{\rm sat}=1/2$. The exchange parameters are set as $J=-2$ and
$J_6=2 J_5$. Spin nematic phase spreads over most of the``non-plateau"
region.
 \label{fig:uuudplateau}}
\end{figure}
Meanwhile, a wide magnetization plateau appears
at $m/m_{\rm s}=1/2$ due to the spin-gapped \emph{uuud} state
in the strong $J_4$ regime~\cite{KuboM,MisguichBLW,MomoiSK}.
Recent magnetization measurement in 2D solid $^3$He
observed a narrow plateau structure at $m/m_{\rm s}=1/2$~\cite{NemaYHI}.
To compare with this, we numerically obtained magnetic phase diagram in $m/m_{\rm s}=1/2$
(Fig.~\ref{fig:uuudplateau}).
The \emph{uuud} phase is easily detected by the four-fold quasi degenerate low-energy states
and a large energy gap above them.
As shown in Fig.~\ref{fig:uuudplateau}, the plateau phase spreads close to the FM phase boundary.
Near its edge, the magnetization process has a very narrow plateau
at $m/m_{\rm s}=1/2$ [see Fig.~\ref{fig:mag_pro}(b)],
which resembles the experimental observation. Even in this case,
spin
nematic phases appear in both low and high magnetization regime.

To summarize, strong competition
between FM two-spin and AF multiple-spin interactions on the
triangular lattice induces the octahedral
spin nematic state with bond antiferro-quadrupolar order.
This state
is a strong candidate for explaining the anomalous magnetic behaviors
experimentally observed in 2D solid $^3$He.

It is our pleasure to acknowledge stimulating discussions with
Hiroshi Fukuyama, H.~Ishimoto, M.~Morishita, H.~Nema, K.~Penc, N.~Shannon,
and R.~Shindou.
Numerical calculations were conducted on RICC in RIKEN
and at IDRIS.
This work was supported by KAKENHI No.\ 17071011, No.\ 22014016, and No.\ 23540397.

\end{document}